\def\mib#1{\mbox{\boldmath $#1$}}

\documentclass[epj]{svjour}
\usepackage[dvips]{graphicx}

\begin{document}
\title{
Analysis of $(\pi^\pm,K^+)$ and $(K^-,K^+)$ hypernuclear production
spectra in distorted wave impulse approximation
}
\author{Hideki Maekawa, Kohsuke Tsubakihara \and Akira Ohnishi
}                     
\offprints{mae@nucl.sci.hokudai.ac.jp} 
\institute{
Department of Physics, Faculty of Science,
Hokkaido University Sapporo 060-0810, Japan
}
\date{Received: date / Revised version: date}
%
\abstract{
We study the hyperon-nucleus potential with distorted wave impulse wave approximation (DWIA)  
using Green's function method.
In order to include the nucleon and hyperon potential effects
in Fermi averaging,  
we introduce the local optimal momentum approximation of target nucleons. 
We can describe the quasi free $\Lambda$, $\Sigma$ and 
$\Xi$ production spectra in a better way than in the 
standard Fermi averaged t-matrix treatments.
} 
\titlerunning{
Analysis of $(\pi^\pm,K^+)$ and $(K^-,K^+)$ hypernuclear production
spectra in DWIA
}
\authorrunning{Maekawa {\em et al.},}
\maketitle
\section{Introduction}
\label{intro}

Study of hyperon-nucleon ($YN$) interaction has an advantage that the 
contributions of meson and quark exchange are different from those in $NN$
interaction, then it may give an opportunity to separate or
 distinguish them.
For example, 
$\Lambda$ does not couple with pions directly
then the strength of the middle range central attraction
would be different in meson and quark exchange pictures.
%
The situation would be clearer for $\Sigma$ hyperons.
Due to the isovector nature of the diquark pair in $\Sigma$,
the Pauli blocking effects between quarks appear in a more direct manner
in $\Sigma{N}$ interaction.
The $\Sigma$ potential in nuclear matter at saturation density
is predicted to be around $+30$ MeV (repulsion)
in a quark cluster model $YN$ potential \cite{Koh1},
while the potential is less repulsive or attractive
in many of the hadronic $YN$ potential models.

Hyperon potential in nuclear matter is also important to understand
compact astrophysical objects such as neutron stars.
The $\Lambda$ hyperon-nucleus potential has been investigated
in the bound region extensively,
and its depth has been known to be about 30 MeV \cite{Mil1}.
For $\Sigma$ hyperon, the bound state spectroscopy is difficult,
because of the strong $\Lambda$ conversion, $\Sigma N\rightarrow \Lambda N$.
In ${}^4_\Sigma$He, which is the only case of observed $\Sigma$
(quasi) bound state \cite{Har1},
the coupling effects is strong and the repulsive contribution
in the $T=$3/2, ${}^3S{}_1$ channel is suppressed,
then it does not strongly constrain the $\Sigma$ potential in nuclear matter.
The analysis of $\Sigma^-$ atomic data suggested
a $\Sigma^-$-nucleus potential having a shallow attractive pocket
around the nuclear surface and repulsion inside the nucleus~\cite{Bat1},
but it is difficult to determine the $\Sigma^-$-nucleus optical potential
in the inner part of nucleus from the atomic data unambiguously.

One of the methods to evaluate the hyperon-nucleus potential is to analyze 
the quasi free (QF) spectrum in the continuum region \cite{Dal1}.
Recent observation of inclusive ($\pi^-$,$K^+$) spectra
on heavy nuclear targets performed at KEK \cite{Nou1}
has made our understanding of $\Sigma$-nucleus potential a step forward. 
In the distorted wave impulse approximation (DWIA) analyses, 
it is suggested that the repulsive real potential of 90 MeV 
or more would be necessary to reproduce the experimental spectra~\cite{Nou1}.
Since this very repulsive $\Sigma$ potential
in nuclei cannot be supported 
by any theoretical models,
it is necessary to verify the validity of approximations 
and prescriptions in the reaction theory currently used for the analysis.
Recently, Harada and Hirabayashi pointed out that on-shell condition
in Fermi averaging (optimal Fermi averaging) for t-matrix of elementary process is important 
to understand the shape of the QF spectrum \cite{Har3},
and their analysis suggests that $\Sigma^-$-nucleus potential
has the repulsive feature in the center of nuclei \cite{Har4}.
A Semi Classical Distorted Wave (SCDW) analysis
by Kohno {\em et al.} \cite{Koh2} also suggests the repulsive nature.
In these works, while the former is based on a fully quantum treatment,
the nucleon and hyperon potential effects are included in the latter.
If the on-shell condition is important
and the difference of the initial (nucleon) and final (hyperon) potentials 
is large,
it would be necessary to take account of 
the effects of the kinematics modification due to the potential energy
in the on-shell condition of the elementary process in nuclear environment
within a quantum mechanical framework
in order to understand the hyperon production spectra
in the QF and bound state region.

In this paper, we investigate the hyperon-nucleus potential
through hyperon production spectra
by introducing the local optimal Fermi averaging t-matrix in DWIA,
which is expected to possess both of the merits in the previous two works.

\section{Model; Green's function method 
and Local optimal Fermi averaged t-matrix}
\label{sec:2}

The Green's function method has been widely applied
to the analysis of hypernuclear reactions. 
This method has the advantage to treat the continuum as well as 
bound state region on the same footing.
In order to include the effects of nucleon Fermi motion
and nucleon/hyperon potentials
into optimal Fermi averaging t-matrix, 
we introduce the Local Optimal Fermi Averaging t-matrix (LOFAt).

Using the Fermi's golden rule,
the differential cross section of $(\pi,K)$ reaction is written as \cite{Aue1},
\begin{equation}
\frac{d^2\sigma}{dE_K d\Omega_K}
=\frac{p_K E_K}{(2\pi\hbar^2)^2 v_\pi}
 \sum_f |T_{fi}|^2\delta(E_\pi+E_T-E_K-E_H)\ ,
\end{equation}
where the subscripts $T$ and $H$ represent target and produced hypernucleus, respectively, and $v_\pi=p_\pi/E_\pi$ is the 
incident particle velocity.

From the angular momentum algebra, we can get the partial wave decomposition of the strength function $S(E)$ 
in the Green's function method \cite{Mor1},
\begin{eqnarray}
\frac{d^2\sigma}{dE_K d\Omega_K}
&=&\frac{p_K E_K}{(2\pi\hbar^2)^2 v_\pi}S(E)\ ,
\\
S(E)&=&\sum_{JM}\sum_{\alpha\beta}\sum_{\alpha'\beta'}
W[\alpha\beta\alpha'\beta']
S_{\alpha\beta\alpha'\beta'}^{JM}(E)\ ,
\\
S_{\alpha\beta\alpha'\beta'}^{JM}(E)&=&
-\frac{1}{\pi}\mbox{Im}\int r^2dr\, r'^2dr'
\tilde{j}^*_{JM}(r) \phi_{\alpha}^*(r) \bar{t}^*(r)
\nonumber\\
&\times&
G^{JM}_{\alpha\beta\alpha'\beta'}(E;r,r')
\bar{t}(r')
\tilde{j}_{JM}(r')
\phi_{\alpha'}(r')\ .
\end{eqnarray}
Here subscripts $\alpha$ and $\beta$ stand for the quantum numbers of
nucleon and hyperon states, respectively.
The coefficient $W[\alpha\beta\alpha'\beta']$ represents
the hypernuclear statistical factor. 
The function $\tilde{j}_{JM}$ is called distorted Bessel function \cite{Tad1}, 
$\phi_{\alpha}(r)$ is the radial wave function of target nucleon,
and $J$ is the total spin of hypernuclei.
The Green's function $G_{\alpha\beta\alpha'\beta'}(E;r,r')$
contains the hypernuclear Hamiltonian $H_H$ then we can get the information of 
optical potential $U_Y$ between hyperon and nucleus.


It was pointed out by Harada and Hirabayashi \cite{Har3} that
on shell kinematics in the Fermi averaging procedure roughly decide
the shape of the QF spectrum
and its prescription of the t-matrix is important. 
We would like to extend their idea by including potential effects. 
Here, we introduce {\em Local Optimal Fermi Averaging} t-matrix (LOFAt),
\begin{equation}
\bar{t}(r;\omega,\mib q) \equiv 
\frac{\int d\mib p_N t(s,t)\rho(p_N)
\delta^4(P_f^\mu(r) - P_i^\mu(r))
}
{\int d\mib p_N \rho(p_N)
\delta^4(P_f^\mu(r) - P_i^\mu(r))
}\ ,
\end{equation}
where
$P^\mu_{i,f}(r)$ denote the total four momenta
in the elementary initial and final two-body states.
We adopt the Fermi distribution function
for the target nucleon momentum distribution 
$\rho(p_N)$ and parameters are taken from \cite{Aue1,All1}.

In obtaining LOFA t-matrix,
we define the nucleon and hyperon energy in nuclei and hypernuclei
containing the nuclear and hypernuclear potential effects,
\begin{equation}
E_B(r)=\sqrt{\mib p_B^2+m_B^2+2m_B V_B(r)}
\sim
m_B+\frac{\mib p_B^2}{2m_B}+V_B(r)
,
\end{equation}
%
where $B=N$ or $Y$.
These treatments enable us to include the potential effects naturally
through the effective mass
$m^{*2} = m_B^2+2m_B V_B(r)$.
Consequently,
the LOFA t-matrix have the dependence on the collision point $r$
through nucleon and hyperon potentials, $V_B(r)$.
It should be noted that the LOFA t-matrix is equivalent to ordinary 
optimal Fermi averaging t-matrix 
when potential effects are switched off.
Product of incoming and outgoing distorted meson waves is evaluated 
in the eikonal approximation. 
In $(\pi^\pm,K^+)$ and $(K^-,K^+)$ reactions at 1.20 GeV/$c$ and 1.65 GeV/$c$,
the isospin averaged cross sections are assumed to be
$\bar{\sigma}_{N\pi^\pm}$=34mb,
$\bar{\sigma}_{NK^+}$=18mb 
and $\bar{\sigma}_{NK^-}$=40mb,
$\bar{\sigma}_{NK^+}$=30mb, respectively.

\section{Results}


\subsection{$\Lambda$ production spectrum}

First, we calculate the $\Lambda$ production spectrum
using the well known parameters from the bound state spectroscopy, 
{\em i.e.} a typical depth of about 30MeV \cite{Mil1},
in order to judge the validity of the present method.

In the calculation, we have assumed
the one body Woods-Saxon type hyperon-nucleus optical potential,
\begin{equation}
U_Y(r)
=(V_{0}^Y+iW_{0}^Y)f(r)
+V_{ls}^Y\frac{\hbar^2 \mib l\cdot\mib s
}{(m_\pi c^2)^2}\frac{1}{r}
\frac{df(r)}{dr}
+V_{C}^Y(r) ,
\end{equation}
with 
$
f(r)=1/(1+e^{\frac{r-R}{d}}),\hspace{5mm}R=r_0(A-1)^{1/3},
$
where $V_{ls}^Y$ and $V_C^Y(r)$ denote the spin-orbit strength
and Coulomb potential, respectively.

\begin{figure}
~~~~~~\resizebox{0.4\textwidth}{!}{%
  \includegraphics[angle=-90]{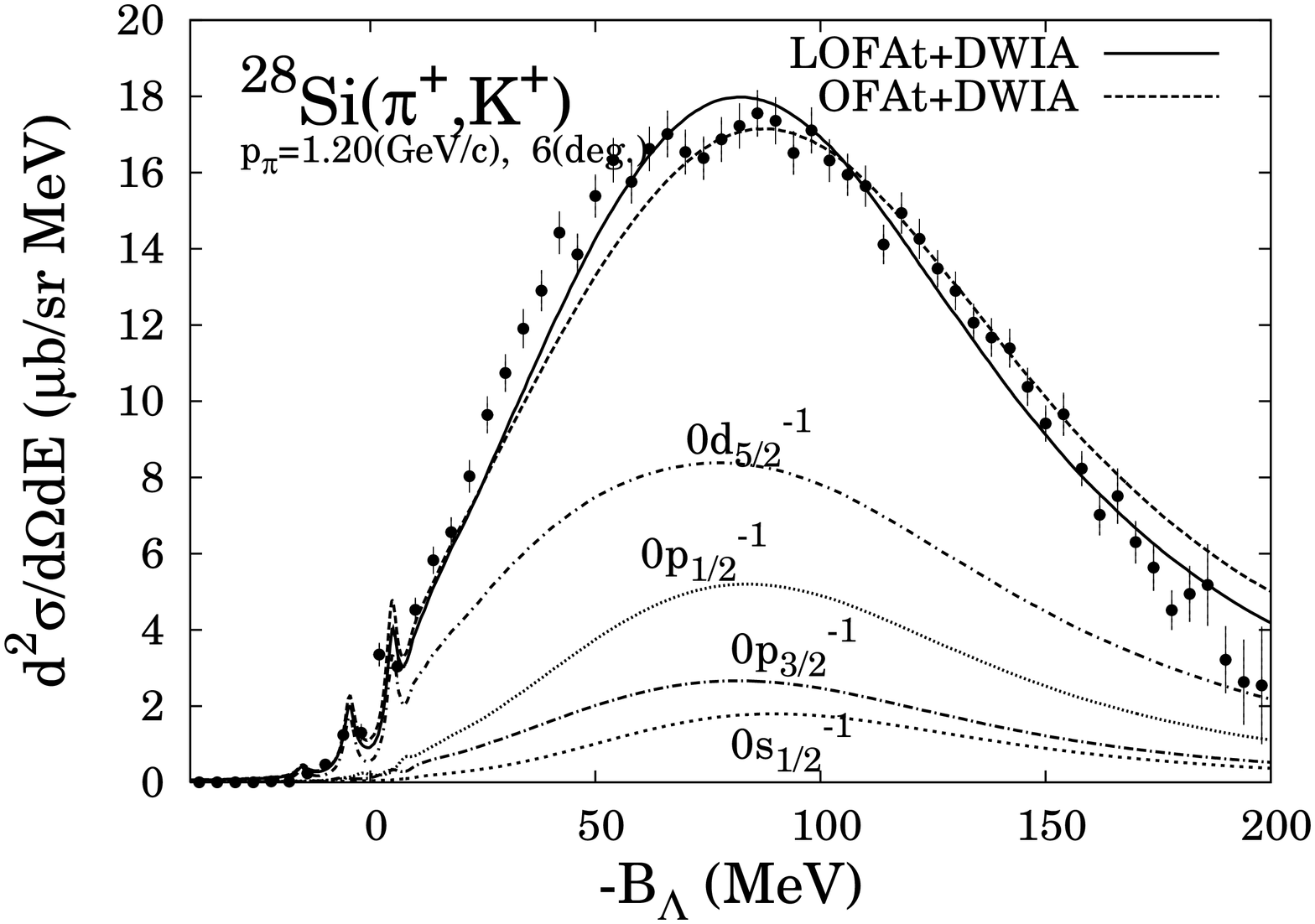}
}\\
~~~~~~\resizebox{0.4\textwidth}{!}{%
  \includegraphics[angle=-90]{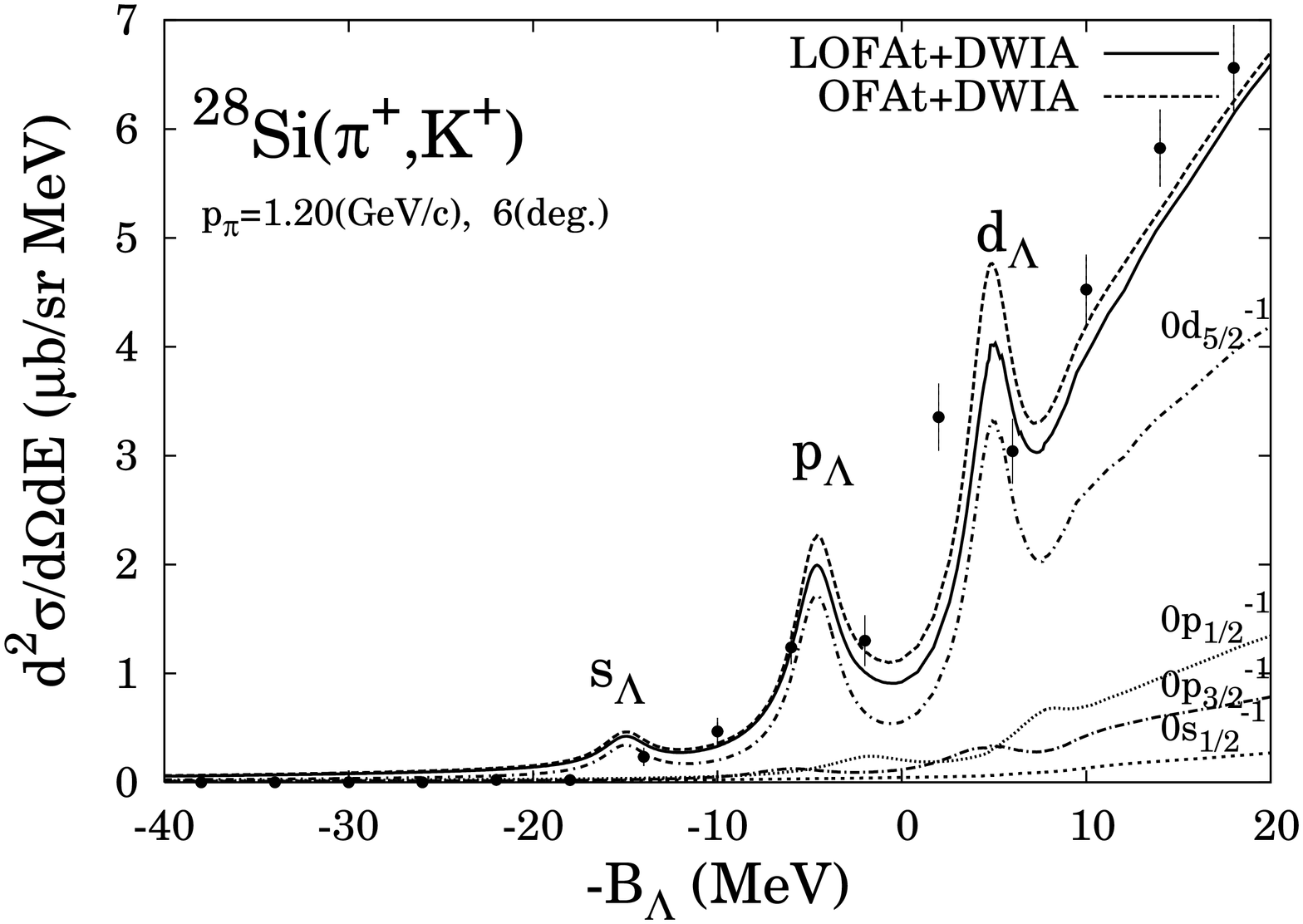}
}
\caption{
\small
The $\Lambda$ hypernuclear production spectrum ${}^{28}$Si$(\pi^+,K^+)$
in the QF region (upper panel) and 
in the bound state region (lower panel)
at $p_\pi$=1.2 GeV/$c$.
Solid line shows LOFAt + DWIA results 
using $\Lambda$-nucleus potential depth of 32 MeV. 
Dotted line shows the Optimal Fermi Averaging t-matrix (OFAt) DWIA result.
Other lines show hole contribution
with 0s1/2, 0p3/2, 0p1/2 and 0d5/2, respectively,
in LOFAt + DWIA.}
\label{Figs:Lambda}
\end{figure}

Figure \ref{Figs:Lambda} shows
the calculated results of $\Lambda$ production spectrum
${}^{28}$Si$(\pi^+,K^+)$ at $p_\pi$=1.20 GeV/$c$, $\theta=6^\circ$ 
in comparison with experimental data.
The experimental data are taken from E438 at KEK.
Solid line shows DWIA results with LOFA t-matrix
with standard parameters
$V_0^\Lambda=-32 \mathrm{MeV}$, $V_{ls}^\Lambda=4 \mathrm{MeV}$,
$r_0=1.1 \mathrm{fm}$ and $d=0.6 \mathrm{fm}$.
We find good agreement of the calculated results with data
in both of QF and bound state regions.

\subsection{$\Sigma^-$ production spectrum}

DWIA analysis in the ordinary on-shell Fermi averaging t-matrix treatment 
can reproduce the $(\pi^-,K^+)$ QF spectrum shape with
the Batty's density dependent (DD) potential and Woods-Saxon potential
with 30 MeV repulsion~\cite{Har3},
but the absolute values are different in these calculations.
It is desirable to describe the spectrum shape as well as the yield,
and the LOFA t-matrix would be helpful for this purpose.

\begin{figure}
~~~~~~\resizebox{0.4\textwidth}{!}{%
  \includegraphics[angle=-90]
{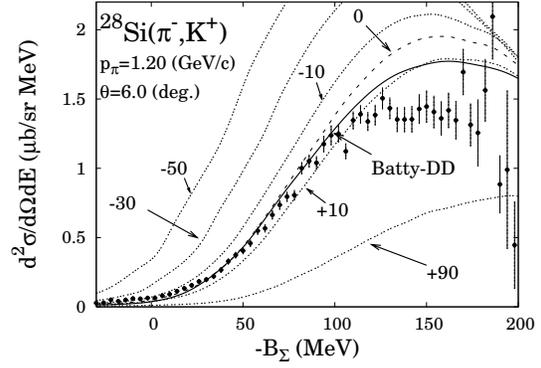}
}
\caption{
\small Differential cross section of $(\pi^-,K^+)$ 
reaction on ${}^{28}$Si target at
the incident momentum of $p_{\pi}$=1.2 GeV/$c$. 
The solid line shows result of Batty's DD potential with LOFAt + DWIA, 
Other line are calculated results with LOFAt + DWIA with potential depth of $V_0$=-50, -30, -10, 0, +10, +90 MeV(up to down), respectively. 
Imaginary part is fixed to be -20 MeV. 
}
\label{Fig:Sigma}
\end{figure}

In Fig.\ref{Fig:Sigma}, we show 
the $\Sigma^-$ production QF spectrum
${}^{28}\mathrm{Si}(\pi^-,K^+)$ at $p_\pi$=1.2 GeV/$c$.
Calculated results using Woods-Saxon type optical potentials
and Batty's DD potential \cite{Bat1}
are compared with experimental data \cite{Nou1}.
It turns out that experimental data on ${}^{28}$Si target is reasonably well
reproduced in Woods-Saxon type potential with small repulsion.
In the Batty's DD potential,
calculated result agrees with the experimental data in 
a wide excitation energy range.
We can see the large potential dependence
in the case of LOFAt + DWIA.

%

\subsection{$\Xi^-$ production spectrum}

\begin{figure}
~~~~\resizebox{0.4\textwidth}{!}{%
  \includegraphics[angle=-90]{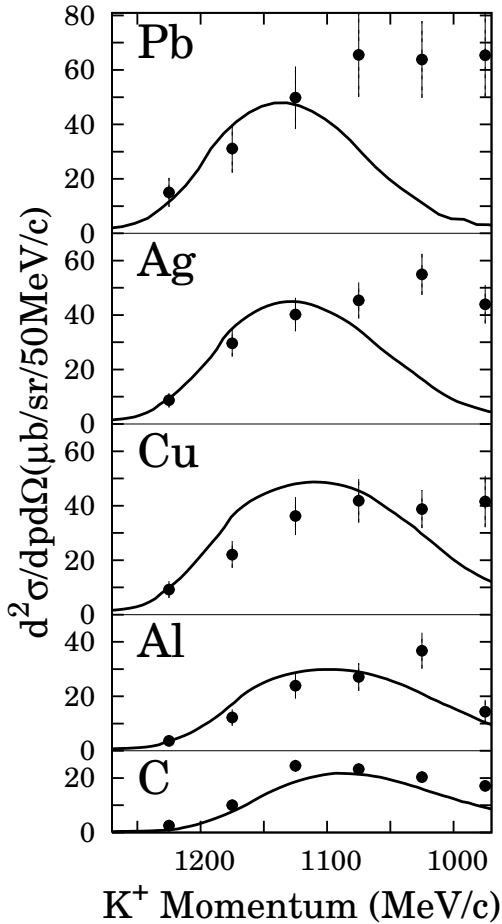}
}
\caption{
\small
The calculated $\Xi^-$-hypernuclear production spectra
in the QF region at $p_\pi$=1.65 GeV/$c$ and 6(deg.)
on C, Al, Cu, Ag and Pb targets in comparison with data~\protect{\cite{Iij1}}.
Solid lines show LOFAt + DWIA results
with $(V^\Xi_0,W^\Xi_0)=(-15 \mathrm{MeV}, -1 \mathrm{MeV})$,
and the experimental resolution is assumed to be 10 MeV.
}
\label{Figs:Xi-QF}
\end{figure}

The depth of the $\Xi^-$-nucleus potential has been suggested to be
around 15 MeV from the analysis of the $(K^-,K^+)$ spectrum
in the bound state region~\cite{Fuk1}.
In that analysis, the observed yield in the bound state region 
is compared with the calculated results,
since the experimental resolution is not enough to distinguish 
the bound state peaks.
Figure \ref{Figs:Xi-QF} shows calculated results of $\Xi^-$ production spectra in LOFAt + DWIA
with potential depth of 15 MeV in comparison with experimental data \cite{Iij1}.
Calculated curves reproduce the experimental data systematically
, while the cross section at lower $p_{K^+}$ region
is underestimated,
where the contribution from multistep processes is important~\cite{Nara}.

\begin{figure}
\resizebox{0.4\textwidth}{!}{%
  \includegraphics[angle=-90]
{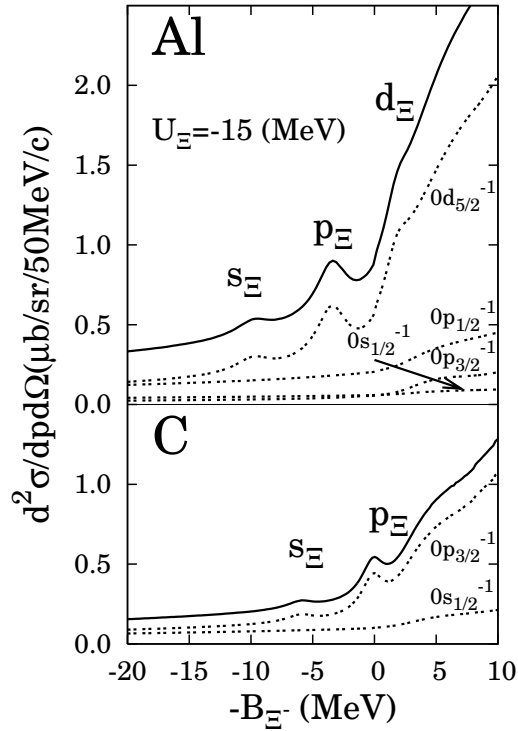}
}
\caption{
\small
The $\Xi^-$-hypernuclear production spectrum in the bound region at $p_\pi$=1.65 GeV/$c$ and 6 (deg.)
on Al (upper panel) and C (lower panel) targets with LOFAt + DWIA. 
Potential parameters are used same as 
Fig. \protect{\ref{Figs:Xi-QF}}.
}
\label{Fig:Xi-BS}
\end{figure}

In Fig. \ref{Fig:Xi-BS},
we show the calculated $K^+$ spectrum in the bound state region
of $(K^-,K^+)$ reactions on ${}^{27}$Al and ${}^{12}$C targets
with the same potential parameters [$(V^\Xi_0,W^\Xi_0)=(-15 \mathrm{MeV}, -1 \mathrm{MeV})$] which explains the QF spectra.
We have assumed an experimental resolution of 2 MeV.
%
We find that bound state peaks are populated selectively
as in the case of $(\pi,K)$ reaction due to high momentum transfer,
and these peaks can be identified if the experimental resolution is
improved to be around 2 MeV.

\section{Summary}

We have studied hyperon-nucleus potentials
through the QF spectra in $(\pi^+,K^+)$ , $(\pi^-,K^+)$ and $(K^-,K^+)$
reactions
using distorted wave impulse approximation (DWIA)
with Local Optimal Fermi Averaging t-matrix (LOFAt) treatment.
In addition to the on shell kinematics~\cite{Har4},
nucleon and hyperon potential effects are included in the Fermi averaging
procedure in LOFAt.
We have found that LOFAt treatment is a better tool
to describe the QF spectrum than standard Fermi averaging prescriptions.
In comparison with the $\Lambda$ production data,
we find good agreement in both of QF and bound state regions
with LOFAt + DWIA.
From the comparison with the $\Sigma^-$ production data,
LOFAt + DWIA result prefers less repulsive $\Sigma^-$ potential than
those suggested in other theoretical models~\cite{Har3,Koh2}.
This difference may come from the kinematics modification 
by the large difference in the initial (nucleon) and final (hyperon)
state potentials.
Finally, we investigate the $\Xi^-$ production spectrum,
and calculated results are found to be in good agreement 
with the experimental QF data using the $\Xi^-$-nucleus potential depth of 15 MeV.
We believe that the present modification would provide a better tool
for the analysis of spectrum in the QF as well as the bound state region.

\section*{Acknowledgements}

We would like to thank Prof. A. Gal, Prof. T. Harada and Prof. M. Kohno
for valuable discussions.
This work is supported in part by the Ministry of Education,
Science, Sports and Culture,
Grant-in-Aid for Scientific Research under the grant numbers,
15540243 and 1707005. 

%

%
%
%
%

\end{document}